# Major Change in Understanding of GRBs at TeV

**Razmik Mirzoyan**[1]

*Max-Planck-Institute for Physics (Werner-Heisenberg-Institute)*
*Foehringer Ring 6, 808805 Munich, Germany*
*E-mail:* `Razmik.Mirzoyan@mpp.mpg.de`

**On behalf of the MAGIC Collaboration**

**Abstract**

Gamma-ray bursts (GRB) are extremely violent, serendipitous sources of electromagnetic radiation in the Universe, occurring at a rate of about once per day. Depending on the emission time two populations of GRBs can be identified; those flaring longer than 2 s are labelled as "long" and the rest as "short". Long-duration GRBs are the most luminous sources of electromagnetic radiation known in the Universe. Their initial prompt flashes of MeV gamma rays are followed by longer-lasting afterglow emission from radio waves to GeV gamma rays. So far the highest energy gamma ray measured from a GRB was a single photon of ~95 GeV, observed by the Fermi-LAT instrument. Emission at TeV energies had been theoretically predicted, but never confirmed by observations. Here we report the detection of a huge signal from GRB 190114C in the TeV energy range by the MAGIC imaging atmospheric Cherenkov telescopes. Starting one minute after the onset of the burst, gamma rays in the energy range 0.2 -1 TeV were observed at more than 50 $\sigma$. This allowed us to study the spectral and temporal development of the GRB, revealing a new emission component in the afterglow with a power comparable to that of the synchrotron component. We found a second peak in the spectral energy distribution of the GRB at an energy of few hundred GeVs. Our modeling, based on the data from the two dozen space- and ground-based instruments that followed GRB 190114C at multiple wavelengths, supports the explanation that the second peak is due to the Inverse Compton radiation mechanism. The two-peaked structure of the spectral energy distribution allows us to constrain some of the key physical parameters of the GRB as the bulk Lorentz factor, minimal electron energy, the ratio of the radiation to magnetic field density. Also the H.E.S.S. imaging atmospheric Cherenkov telescope recently reported on a 5 $\sigma$ gamma-ray signal from the GRB 180720B, measured in the afterglow phase, 10 hours after the onset of the explosion. These observations prove that the GRBs are more powerful than assumed until recently. Because the observed GRBs did not show peculiar properties, we believe that from now on detection of gamma-ray signal from GRB afterglows at very high energies will become one of the standard observations.

[1]Speaker

## 1. Introduction

GRBs were discovered in late 1960's as serendipitous sources of extremely intense MeV gamma rays. The first publication goes back to 1973 [1]. GRBs appear at random locations and times in the sky, at a rate of about once per day. For a short moment they become the brightest sources in the sky but then rapidly fade away. Over the past ~50 years we learned a lot about their nature [2-5]. Interestingly, it took quite some time to find out that GRBs occur at cosmological distances and that these are the most luminous sources of electromagnetic emission known in the Universe [6].

Depending on the duration of the main emission GRBs can be classified as being of "short" or "long" type. The border line between the two populations is 2s. Long GRBs are initiated in the cores of some dying massive stars undergoing gravitational collapse [7]; such evidence is supported by supernovae occurring in coincidence. Short GRBs are likely triggered by the mergers of binary neutron star systems. The gravitational wave event GW170817 supported this long-standing hypothesis by appearing also as GRB 170817A [8].

The initial phase of a long GRB can last up to hundreds of seconds and is characterized by a "prompt" emission, mostly at MeV energies. One believes that under certain conditions the massive star collapse launches collimated jets of plasma, which expand with ultra-relativistic velocities [2,3]. The prompt emission originates from within the jets. The observed sub-millisecond variability timescales and associated high-energy photons support this assumption.

The "afterglow" follows the prompt phase. It is characterized by emission over a very broad wavelength range and smooth decay over much longer timescales. It is widely accepted that the so-called external shocks, produced when the jets pierce the ambient gas, are responsible for the afterglow.

Despite the progress, still many basic questions about GRBs as, for example, the acceleration mechanisms of particles, which produce the prompt and the afterglow emissions, or formation of the jets, remain open [5].
Afterglow emission was known to extend from radio frequencies up to GeV energies. It is believed to be mainly due to synchrotron radiation from energetic electrons that are accelerated within magnetized plasma at the external shocks [2,3].
The synchrotron radiation from energetic electrons within the relativistic expanding jet could be responsible for the prompt emission, but still this needs to be explored further.

Till the recent detection of TeV gamma rays from GRBs, shown in this report, so far the highest energy record-holder was a single photon of ~ 95 GeV. The latter was measured by the Large Area Telescope (LAT) onboard the Fermi Gamma-ray Space Telescope (hereafter Fermi) from GRB 130427A at redshift $z \sim 0.34$ [9].
It is interesting to note that such an energetic photon is close to the maximum energy that, according to theory, the synchrotron process can produce in the afterglow; see §4 below for more details.

GRB emission was anticipated also in the much higher, so-called very high energy (VHE) domain due to, for example, the inverse Compton (IC) process.
Over the past 2-3 decades, numerous searches for TeV emission from GRBs have been carried out by using diverse instruments and observational techniques, but none of them succeeded [11,12]. One can find only a couple of weak or ambiguous hints in the literature; these will be discussed in the text below.

In the IC mechanism the population of energetic electrons, which produces the synchrotron radiation, scatter on the low-energy photons and boost their energy into the VHE range [13-17,46]. Along with synchrotron emission, IC is regularly observed from, for example, the blazar sub-class of active galactic nuclei. Sometimes IC can become the dominant emission mechanism of a source [18,19].

For observing a GRB from the very first moment one needs a telescope with a wide field of view (FoV). Also a narrow FoV telescope has a chance to observe the GRB albeit not from the very first moment. For this it needs to receive a fast alert from an external wide FoV instrument.

## 2. Detection of GRB 190114C

GRB 190114C was first identified as a long-duration GRB by the BAT instrument onboard the Neil Gehrels Swift Observatory (Swift) [20] and the Gamma-ray Burst Monitor (GBM) instrument onboard the Fermi satellite [21] on 14th January 2019, 20:57:03 Universal Time (UT) (hereafter T0). The T90 (time interval containing 90% of the total photon counts) was measured to be 116 s by Fermi/GBM [21] and 362 s by Swift/BAT [22]. Very soon other reports followed. Triggered by the Swift/BAT alert, the Major Atmospheric Gamma Imaging Cherenkov (MAGIC) telescopes observed GRB 190114C from T0 +57 seconds until T0 +15912 seconds. Based on online analysis MAGIC reported detection of gamma rays above 0.3 TeV with more than 20 σ significance for the first 20 minutes of data. About 4 hours after the burst alert also MAGIC issued an Astronomy Telegram (ATel #12390) [23] and a GCN Circular (# 23701) [24] on the first detection of TeV gamma rays from a GRB.

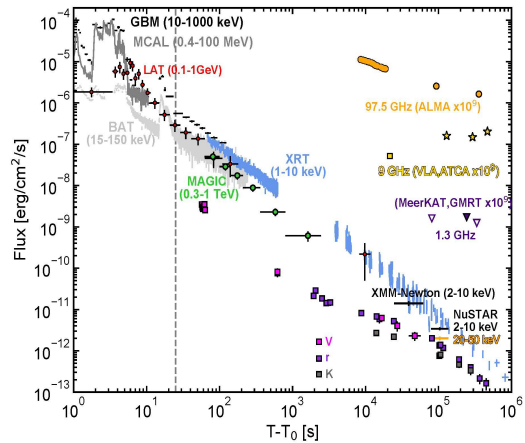

Figure 1. Energy flux development in time for GRB 190114C as measured by two dozen space-born and ground-based instruments, covering the range from ~ 1.3 GHz to ~1 TeV; see details in [25].

These sparked big interest and about two dozen space born and ground-based instruments observed GRB 190114C at different wavelengths, from 1.3 GHz and up to ~1 TeV, see the results on Fig. 1. The details can be found in [25].

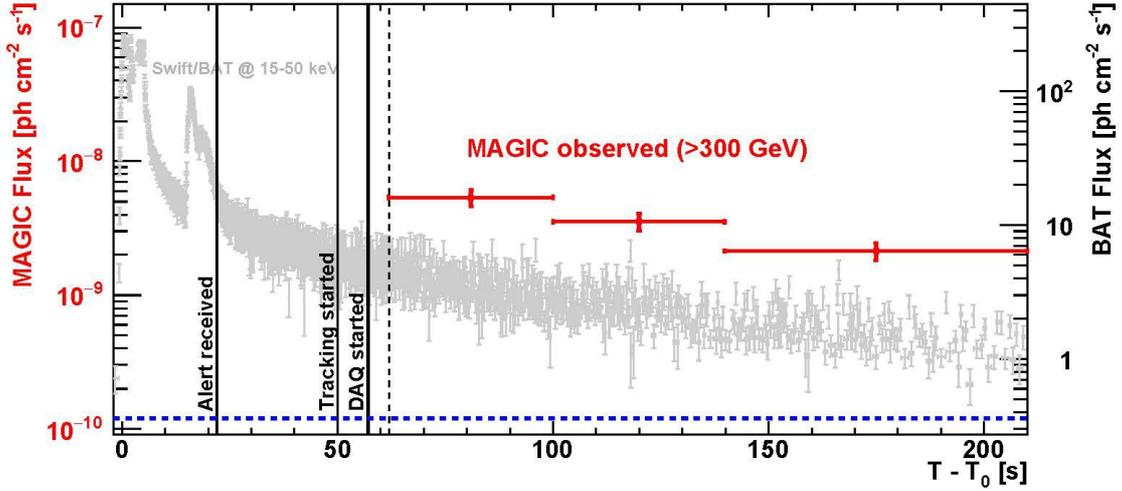

Figure 2. Light curves by MAGIC (≥ 0.3 TeV) and by Swift-BAT (15-50 keV) for GRB 190114C for the first 210 seconds. The horizontal blue-dashed line shows the MAGIC flux for Crab Nebula for E ≥ 0.3 TeV. The solid vertical lines (1-3, from left) show the time sequence of actions for MAGIC. The vertical dashed line shows the start time of the DAQ system under stable conditions. Image taken from [26].

The subsequent offline analysis of the collected data showed an extremely strong detection with significance > 51 $\sigma$, see Fig. 3 [26]. Also the measurement of its redshift z = 0.42 has been reported, see [26] and references therein. The isotropic-equivalent energy of the emission in the interval 10–1000 keV observed by Fermi/GBM at T90 was estimated to be $E_{iso}$ = 3 x $10^{53}$ erg. This tells that GRB 190114C was a rather energetic, but not an exceptional burst.

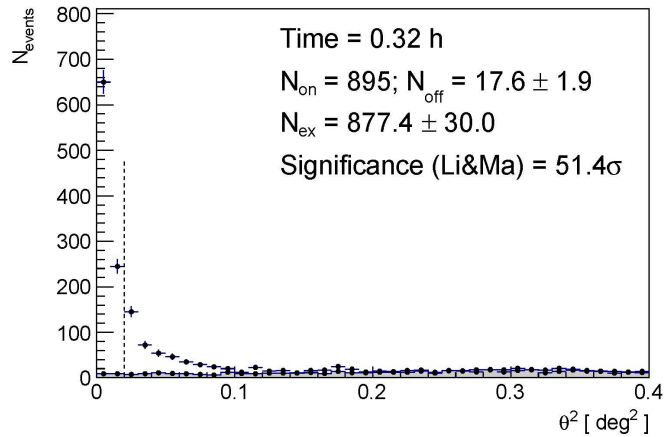

Figure 3. Significance of the $\gamma$-ray signal between T0 + 62 s and T0 + 1,227 s measured by MAGIC from GRB 190114C. Image taken from [26].

On Fig. 3 we show the distribution of the squared angular distance, $\theta^2$, for the gamma-ray signal (points) and background events (grey shaded area). The dashed vertical line shows the applied $\theta^2$ cut for the signal and the background regions. The significance is calculated using the Li & Ma method [27]. The significance for energies ≥ 0.7 TeV is 5.8 $\sigma$ and for energies ≥ 1.1

TeV it is 2.5 σ [26]. This is the most intense signal ever measured from a gamma-ray source since the development of the ground-based VHE gamma-ray astronomy, in the first 30 s of observations the measured rate of gamma rays from GRB 190114C was ~130 times higher than that from Crab Nebula, the strongest source and the standard candle for VHE in our Galaxy.

The observed gamma-ray spectra from cosmologically distant sources gradually change their shape, becoming steeper at high energies above few tens of GeV. This flux attenuation, which is stronger the higher is the energy of the photon and the further is the source, is due to its interaction with the low energy photons $E_{EBL}$ of the extra-galactic background light (EBL). The latter is a diffuse, abundant background of infrared, optical, and ultraviolet radiation emitted by all stars and galaxies during the existence of the Universe, which fill in the space. As a result of such interaction, under the condition that the center of mass energy condition $\sqrt{E_\gamma E_{EBL}} \geq 2m_e c^2$ can be fulfilled, an $e^-e^+$ pair is born. The gamma-ray spectrum in close vicinity of sources, where the EBL attenuation plays no role, can be referred to as the source spectrum. Taking as a basis a plausible model of the EBL [28], one can infer the source spectrum by deconvolving the measured spectrum from the anticipated attenuation.

Because the MAGIC observation started about one minute after the onset of the GRB 190114C (see Fig. 1), the reported TeV emission should be associated with the afterglow phase although a partial overlap with the prompt emission phase cannot be excluded [26].

**2.1. The MAGIC Telescopes Pioneered Measurements down to few tens of GeV**

Since 2009 the MAGIC telescope project operates two 17-m diameter IACTs (MAGIC-I and MAGIC-II) in a coincidence (stereoscopic) mode at the *Roque de los Muchachos* European Northern Observatory in La Palma, Canary Islands, Spain [29, 30]. By imaging the Cherenkov light emission from extended air showers within a field of view of ~10 square degrees, the MAGIC telescopes can very efficiently discriminate the vast majority of isotropic hadron background and detect gamma rays above an energy threshold of 30 GeV (Sum-Trigger-2, see §2.4) or alternatively 50 GeV (Standard Trigger). The MAGIC-I telescope project was initiated in 1995 as a stand-alone telescope [31]. It had a pioneering design to become the first instrument measuring in the sub-200 GeV energy domain, down to ~10 GeV. It is interesting to note that back in mid 1990's the young community of IACTs believed that it was impossible to lower the threshold of the air Cherenkov technique much below ~300 GeV. The reason was the widely accepted belief that the threshold of an IACT, similar with non-imaging Cherenkov detectors, is defined by the fluctuations of the Light of the Night Sky (LoNS) (see, for example, the review [32]). In mid-1990's the initiators of the MAGIC project understood that the aforementioned assumption was not correct and in fact a large-size IACT of fast electro-optical design can successfully operate above a threshold of few tens of GeV [33]. Several innovative design features were put forward for MAGIC for achieving the goal of very low energy threshold (use of optimized light guides, high-efficiency, nanosecond fast novel hemi-spherical PMTs and analog signal transmission via optical fibers, GSample/s fast signal digitization, parabolic-shape reflector, etc).

One of the main design goals was the repositioning of the telescope within ~20 s to an arbitrary position in the sky for measuring a possible signal from a GRB. For that purpose the reflector frame was made from light-weight reinforced carbon-fiber and the novel light-weight individual mirror segments were set on stepping motor driven actuators; these are operated by using the Active Mirror Control (AMC) system. After many improvements over years, today the light-weight telescopes (~ 70 T) can be re-positioned to an arbitrary position in the sky within 25 s.

MAGIC, as well as the other IACTs, are tuned to follow external alerts provided by satellite instruments with large FoV; some of those provide precise coordinates of the triggered GRBs within several seconds.

**2.2. TeV-band observations of GRBs with MAGIC and other facilities in the past**

The search for TeV gamma rays from GRBs has been pursued over many years by employing diverse experimental techniques, but no clear detection was reported, see [26] and the references therein.

MAGIC is responding to GRB alerts since 15th July 2004. For the first 5 years, MAGIC was operated as a single telescope (MAGIC-I). After the second telescope (MAGIC-II) was added in 2009, GRB observations have been carried out in the coincidence (stereoscopic) mode. MAGIC observed 105 GRBs from July 2004 to February 2019. Of these, 40 have defined redshifts. 8 GRBs had a redshift below 1 and 3 GRBs below 0.5. Observations started less than 30 minutes after the burst for 66 events (of which 33 lack the redshift), and less than 60 seconds for 14 events. The small number in the latter case is mainly due to adverse weather conditions at the time of the alerts.

Despite 15 years of dedicated efforts, no unambiguous evidence for gamma-ray signals from GRBs could be detected by MAGIC before GRB 190114C. The flux upper limits for GRBs observed in 2005-2006 were found to be consistent with simple power-law extrapolations of their low-energy spectra. More detailed studies were presented for GRB 080430 and GRB 090102 that were simultaneously observed with MAGIC and other instruments in different energy bands. Since 2013, GRB observations are performed with the below described automatic procedure. In addition, for some bright GRBs detected by Fermi/LAT, late-time observations have been conducted, up to one day after the burst for searching for potential emission in the afterglow phase.

Observation of GRB 190114C can be compared with those of other GRBs followed by MAGIC under similar conditions. The chance to detect a GRB by an IACT depends on the source redshift z, the zenith angle (higher energy threshold for larger zenith angles) and ambient lighting conditions. Only four GRBs were observed under the conditions $z < 1$ and $T_{delay} < 1$ h (Table 1). Except for GRB 190114C, the rest are short GRBs. This is not surprising because it is known that compared to long GRBs the redshift distribution for short ones is shifted to lower values. Few more long GRBs with $z < 1$ were actually followed up by MAGIC with $T_{delay} < 1$ h, but the observations were not successful due to technical problems or adverse observing conditions. More details can be found in [26].

Interestingly, the measured by MAGIC short GRB 160821B was nearby, it's redshift was 0.16. MAGIC reached the alerted source position in 24 s and continued observations for ~4 h. The observing weather conditions were not optimal and there was a bright Moon in the sky. Three independent analyses were performed. These, after including the number of trials, converged on 3.1 σ signal for energies ≥ 500 GeV [35]. Diverse test results support the evidence of a genuine signal but the low signal strength does not allow making a serious claim.

Below we discuss briefly a few ambiguous hints, which one can find in the literature.

The AIROBICC cosmic and gamma-ray detector was part of the HEGRA array. It measured a small, spurious excess from the GRB 920925C for energies ≥ 20 TeV, with a significance of 2.7 σ after trials [36]. Note that the location of the highest excess was at 9° distance from the probable target position, measured by the EURECA/WATCH hard X-ray mission. Moreover, keeping in mind the cited very high energy threshold and the strong EBL absorption for the GRB at a possibly cosmological distance scale, it is highly improbable that the reported excess was due to a GRB.

Milagrito, the smaller version of the Milagro, the predecessor of HAWC, was a very small size (1700 m²) detector. It measured a spurious signal from GRB 970417A. The after trial probability of this event was estimated to be $1.5 \times 10^{-3}$ for energies ≥ 650 GeV. The red shift of the source was unknown. The estimated spectral energy distribution of this event looked unusual because the fluence for energies ≥ 650 GeV was higher than that measured by BATSE by more than one order of magnitude. To "rescue" this unusual situation one may assume that the peak in the X-ray range is at much lower energies than what has been measured, but this is improbable. The authors of the paper [37] skeptically mentioned that possibility.

Short summaries of the GRB observations by Veritas, H.E.S.S., HAWC, Milagrito and Milagro in the past can be found in [26]. Unfortunately none of those few hundred observations showed any convincing signal.

| Event | redshift | $T_{delay}$ (s) | Zenith angle (deg) |
|---|---|---|---|
| GRB 061217 | 0.83 | 786.0 | 59.9 |
| GRB 100816A | 0.80 | 1439.0 | 26.0 |
| GRB 160821B | 0.16 | 24.0 | 34.0 |
| GRB 190114C | 0.42 | 58.0 | 55.8 |

Table 1. List of selected GRBs (z < 1 and $t_{delay}$ < 1h) observed by MAGIC under good conditions. Only the GRB 190114C is of long type.

**2.3. The Automatic Alert system**

An Automatic Alert System (AAS) has been developed with the aim of the fastest possible response to GRB alerts. It is a multi-threaded program that connects to the Gamma-ray Coordinates Network (GCN) servers, receives notices with sky coordinates of GRBs, and sends commands to the Central Control (CC) software of the MAGIC telescopes. This includes a check of the visibility of the target according to predefined criteria. A priority list was set up for the case when several different types of alerts are received quasi-simultaneously. Moreover, for the case of multiple alerts for the same GRB, the AAS will select the one with the best localization.

When an alert is evaluated as observable by the AAS, the telescopes will be automatically slewed to the target position in the sky. An automatic procedure, implemented in 2013, prepares the subsystems for data taking during the telescope slewing; data taken immediately before the alert will be saved, relevant trigger tables will be loaded, the individual electronic trigger thresholds will be set and the AMC system will adjust the individual mirrors to provide the best parabolic shape at the target position [38]. While moving, the imaging cameras continue the calibration procedure. The Data Acquisition (DAQ) system continues taking data while it receives information about the target from the Central Control software. The rate limiter will be set to 1 kHz for preventing from extremely high rates and for avoiding the saturation of the DAQ system. When the repositioning completes, the target is tracked in wobble mode, the standard observing mode for MAGIC. To date, the fastest GRB follow-up was achieved for the GRB 160821B, when the data taking started only 24 seconds after receiving the alert.

### 2.4. The operational very low energy threshold of MAGIC and its readiness to promptly respond to alerts for measuring GRBs

The goal to measure distant extragalactic sources and GRBs along with pulsars guided the the design of the MAGIC telescope. Today we can operate the telescopes with the *standard trigger*, providing a threshold of ~50 GeV and with the Sum-Trigger, providing a threshold of ~30 GeV. Already in 2008, by using the first version of Sum-Trigger, we measured pulsed gamma rays from the Crab pulsar above the energy threshold of 25 GeV with the stand-alone MAGIC-I [39]. The improved Sum-Trigger-2 system, implemented in both telescopes, allows us to measure the Crab pulsar above the threshold of ~20 GeV with 5 $\sigma$ signal in less than 6 hours (the quality factor $Q=2.1\cdot\sqrt{t}$). The steady emission of the Crab Nebula can be measured above the energy threshold of ~30 GeV [40]. The spectrum of the Geminga pulsar was measured above the energy threshold of 16 GeV [41].

Results of those measurements strengthened our fidelity that the follow-up of appropriate GRB candidates will lead MAGIC to a successful measurement above the energy threshold of few tens of GeV and it should be just a matter of time to find the first signal.

For "polishing" the complex, fully automatic procedure used to observe GRBs, once every month, during the data taking shift in La Palma we issue a fake GRB alert; only a few persons know about the origin of the trigger. Typically the fake location is chosen to be close to the one of the known intense gamma-ray emitters. In this way we are able to check and further improve the performance of diverse subsystems as well as of the entire system of the telescopes.

### 2.5. Operating MAGICs in the presence of partial moonlight, at dusk and dawn

Already in mid 1990's the researchers from the HEGRA collaboration developed the IACT observation technique in the presence of partial moonlight [42]. MAGIC adopted and further improved this technique, performing regular observations at dusk, dawn and partial moonlight [43]. The use of this technique allows one to prolong by ~30 % the IACT observations of celestial sources albeit, because of the higher noise, at the expense of somewhat higher energy threshold.

### 2.6. The surprise of detecting the GRB 190114C marked the beginning of a new era in the understanding of GRBs

Due to upgrade of the telescopes in 2013 and a number of following improvements MAGIC has arrived at its best sensitivity of ~0.6 % Crab [30]. The sensitivity has been further boosted at the lowest energies due to the use of the Sum-Trigger-2 and at the highest energies (~100 TeV) due to the use of the so-called very large zenith angle observation technique; observation of sources in the zenith angle range 70° - 80° can provide a shower collection area on the order of 1-2 km² [44,45].
The telescopes can observe sources starting from few tens of GeV at low zenith angles and up to ~100 TeV at the zenith angle of 80°.

While we were anticipating to measure a gamma-ray signal from the next best remote GRB candidate above a very low energy threshold, we measured instead a gigantic signal from the GRB 190114C at TeV energies, in the presence of partial moonlight and from the zenith angle range ≥ 60°. This detection marked a new era in the understanding of GRBs. According to Monte Carlo simulations we measured a signal from GRB 190114C above the energy threshold of ~200 GeV; the relatively high threshold was due to the presence of partial Moon and the large zenith angle.

### 2.7. As to why GRB 190114C happened during moonlight and at a large zenith angle

The number of GRB candidates that an IACT can follow is proportional to the solid angle it can cover in the sky. Therefore it is much more probable that a GRB will appear in the field of view of an IACT at a large zenith angle.
Because the energy threshold of an IACT increases with observation zenith angle, it is more probable to measure a GRB at higher energies rather than close to its energy threshold.
The probability of a GRB to happen in the field of view of a telescope is proportional to its operational time. Typically the IACTs are operating at dark nights with a clear sky, which makes only ~10% from the yearly available time. Operating an IACT at partial Moon and dusk and dawn increase this resource by another ~30 %, thus increasing the probability to detect a GRB.

### 3. MAGIC observations of GRB 190114C.

On the night of 14 January 2019, at 20:57:25 UT (T0+22 s), Swift/BAT distributed an alert reporting the first estimated coordinates of GRB 190114C (RA: +03h 38m 02s; Dec: -26d 56m 18s). The AAS validated it as observable and triggered the automatic re-pointing procedure, and the telescopes began slewing in fast mode from the position before the alert. The MAGIC-I and MAGIC-II telescopes arrived on the target and began tracking GRB190114C at 20:57:53 UT (T0 + 50 s). After starting the slewing, the telescopes reached the target position in approximately 27 seconds. At the end of the slewing, the cameras on the telescopes oscillated for a short time. Later on we reproduced the exact motion of the telescopes, and measured the oscillations of the imaging cameras by using four high-rate CCD and CMOS cameras, watching the imaging camera behavior of both telescopes. We verified that the duration of the oscillations was less than 2 seconds after the start of tracking, and its amplitude was below 1mm (0.6 arc-minutes) when data taking started. Data acquisition started at 20:58:00 (T0 + 57 s) and the DAQ system was operating stably from 20:58:05 (T0 + 62 s), as shown in Fig. 2. Observations were performed in the presence of moonlight, approximately 6 times the level for dark observations. Data taking for GRB190114C stopped on 15 January 2019, 01:22:15 UT. The total exposure time for GRB190114C was 4.12 h.

### 3.1. MAGIC data analysis for GRB 190114C

Data collected from GRB190114C was analyzed by using the standard MAGIC analysis software [30] and the analysis chain tuned for data taken under moonlight conditions [43]. No detailed information on the atmospheric transmission was available since the micro-LIDAR was not operating during the night of the observation. Because of that the quality of the data was assessed by checking the other auxiliary weather monitoring devices such as the star-guider camera, the heat-sensing infrared camera for the presence of clouds as well as the stability of the DAQ rates.

A dedicated set of Monte Carlo (MC) simulation gamma-ray data was produced for the analysis, matching the trigger settings (discriminator thresholds), the zenith-azimuth distribution, and the LoNS level during the GRB 190114C observations; details can be found in [26]. The final data-set comprises events starting from 20:58:05 UT. Due to the higher level of LoNS, compared to standard analysis, a higher level of image cleaning was applied to both the real and the MC data [30]. A higher threshold cut on the integrated charge of the event image, set to 80 photo-electrons, was used for evaluating the photon flux. Details of the analyses, including the assessment of the absolute energy scale as well as the small impact of uncertainties of tested several EBL models can be found in [26]. The spectra in Figure 4 were derived by assuming a simple power law function for the source intrinsic spectrum.

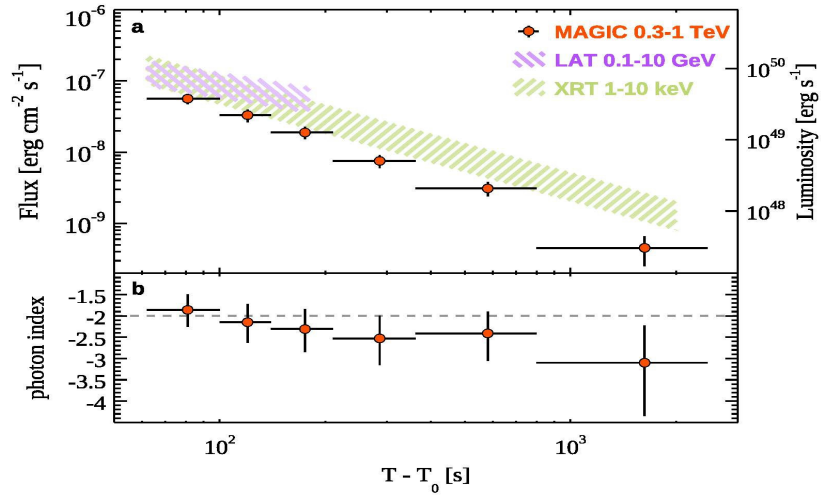

Figure 4. Light curves in KeV, GeV and TeV, and spectral evolution in the TeV band for GRB 190114C. **a**, Light curves in energy flux (left axis) and apparent luminosity (right axis), for MAGIC at 0.3–1 TeV (red symbols), Fermi-LAT at 0.1–10 GeV (purple) and Swift XRT at 1–10 keV (green). For the MAGIC data the source flux is corrected for EBL [26]. **b**, Power-law photon index is taken from time-resolved intrinsic spectra. The horizontal dashed line shows the value −2. Shown errors are statistical, 1 σ [26].

The light curve for the EBL corrected MAGIC flux in the range 0.3-1 TeV can be well fitted by a simple power-low function $F(t) \approx t^\beta$ with $\beta = -1.60 \pm 0.07$.

The striking feature seen on Fig. 1 and Fig. 4 is that the temporal decays of the energy flux in the TeV and X-ray bands show similar behavior. One may assume that the VHE emission is correlated with the electron synchrotron emission. This hints to a "leptonic" emission mechanism. The well-known inverse Compton radiation, where the energetic electrons in the external shock scatter on ambient low-energy photons, boosting these to VHE energies, could offer a viable scenario.

If the acceleration of electrons and protons in the external shock occurs in a correlated manner, also the ultra-high energy protons could offer an explanation for the observed spectra. But the low radiative efficiency of hadron-induced processes does not support such an assumption.

On Fig. 5 one can see the observed (gray open circles) and the derived source spectra (blue filled circles) after unfolding and correcting for the EBL absorption for "true" energies E ≥ 0.2 TeV, averaged over the initial ~40 minutes of observation. The errors on the flux correspond to 1 σ. The upper limits at 95% confidence level are shown for the first non-significant bin at high energies. Also shown is the best-fit model for the source spectrum (black curve) when assuming a power-law function. The grey solid curve for the observed spectrum is a convolution of this curve with EBL. The grey dashed curve is the forward-folding fit to the observed spectrum with a power-law function. The observed spectrum can be fitted with a power law with an exponent $\alpha_{obs}=-5.43\pm0.22$ (only the statistical error). Obviously the "culprit" for such a very steep spectrum is the strong absorption by the EBL. By using the plausible EBL models we estimated, for example, that the gamma-ray flux of 1 TeV photons is attenuated 300 times. Nevertheless, because of the extremely strong signal we could observe γ-rays of energy higher than 1 TeV. By de-convoluting the measured spectrum from the absorption effect of the EBL by using the plausible model [28], we obtained the GRB spectrum, which can be well described by a power law with $\alpha_{source} =-2.22$ (+0.23 -0.25) (statistical errors only). Use of other EBL models showed that the spectral index changes within the cited above errors [26]. Interestingly, we could not observe any cut-off or break in the spectrum at the highest energies; with confidence level of 95 % it extends beyond 1 TeV. The fact that the spectrum can be described by a power law of ~2 tells that there is equal power radiated in the measured energy range 0.2-1 TeV and possibly beyond. This ascertains that a significant fraction of the GRB radiation is emitted in the TeV energy range.

**4. Synchrotron burnoff limit for the afterglow emission**

The expanding relativistic jets from GRBs decelerate and dissipate their kinetic energy in the ambient medium, producing shocks, where electrons can be accelerated [5]. The origin of afterglow is the radiation of those electrons. The maximum energy of electrons that can be attained in the reference frame co-moving with the post-shock region can be estimated by equating the timescales of acceleration and energy loss, the latter primarily due to synchrotron emission [10]. It can be shown that the maximum energy of synchrotron emission is independent of *B* see, for example, [26].

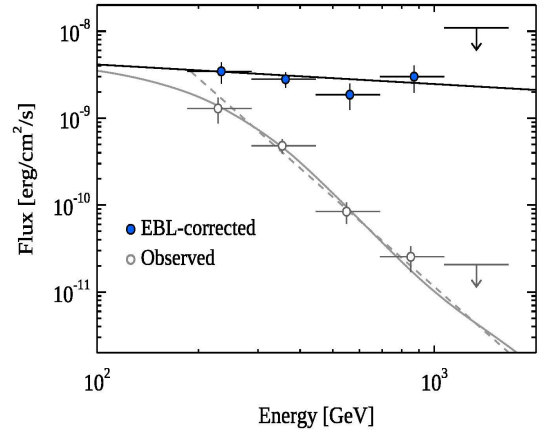

Figure 5. MAGIC spectrum above 0.2 TeV averaged over the period between T0 + 62 s and T0 + 2,454 s for GRB 190114C. Image taken from [26].

The observed spectrum of afterglow synchrotron emission is expected to display a cutoff below the energy $\varepsilon_{syn;max} \approx$ 100 MeV x *[$\Gamma_b(t)/(1 + z)$]*, which depends only on the time-dependent bulk Lorentz factor *$\Gamma_b(t)$* of the external shock. The latter can be derived from the solution of the dynamical equations of the external shock [12], which allows one to estimate $\varepsilon_{syn;max}$ and its evolution.

It can be shown that even under extreme assumptions, resulting in very high values of $\varepsilon_{syn;max}$, it is still well below the energy of photons detected by MAGIC; see for details [26].

Also synchrotron emission by protons accelerated to ultrahigh-energies in the external shock has been proposed as a mechanism for GeV-TeV emission in GRB afterglows, potentially at energies above the burn-off limit for electron synchrotron emission, see for details [26] and references therein.

Although proton synchrotron emission may possibly explain the GeV emission observed in some GRBs, due to its low radiative efficiency, it is strongly disfavored as the origin of the luminous TeV emission observed in GRB 190114C [26]. A more plausible mechanism may be inverse Compton emission by accelerated electrons, see [25,26] and references therein.

## 5. Spectral and temporal developments of GRB 190114C from observations at multiple wavelengths

The extremely strong signal measured by MAGIC allowed us to follow both the temporal and spectral developments of the GRB 190114C.
The spectral energy distributions (SEDs) of the radiation detected by MAGIC in five consecutive time intervals are shown in Fig. 6. For the first two intervals also the observations in the GeV and X-ray bands are available. During the first interval of 42 s duration (68–110 s; blue data points and blue confidence regions), Swift-XRT, Swift-BAT and Fermi-GBM data show that the synchrotron component peaks in the X-ray band. At higher energies, up to 1 GeV, the SED is a decreasing function of energy (Fermi-LAT data at 0.1-0.4 GeV). On the contrary, the MAGIC flux at much higher energies ≥ 0.2 TeV shows a harder spectrum. This evidence, independent of the chosen EBL model [25], shows that the TeV radiation cannot be a mere higher energy extension of the known afterglow synchrotron emission; it is a new spectral component. The extended duration and the smooth, power-law temporal decay of the radiation detected by MAGIC (see Fig. 1 and Fig. 4) suggest a correlation between the TeV emission and the broadband afterglow emission [25,26]. As already mentioned, the synchrotron self-Compton (SSC) radiation in the external forward shock can offer a natural explanation. In that scenario a second spectral component, peaking at very high energies, is produced.

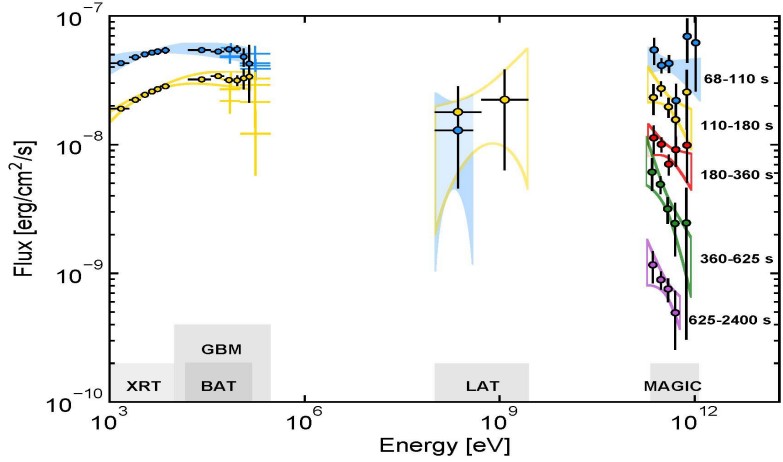

Figure 6. SED development of GRB 190114C in time measured by Swift XRT/BAT, Fermi GBM/LAT and by MAGIC. The MAGIC measurement is split into 5 consecutive periods of time. For the first 2 time bins also the data from the above-mentioned satellites are shown. Image taken from [25].

Though the SSC emission has been predicted for GRB afterglows, the forecasted spectrum and luminosity suffered from large uncertainties of poorly known physical parameters in the emission region, as, for example, the magnetic field strength. For the first time the strong detection of MAGIC made it possible to infer the important physical parameters for adopting the SSC model.
Our studies show that with the development of the GRB afterglow the SSC peak moves to lower energies (see the development of the MAGIC SED in consecutive time bins on Fig. 6) and crosses the MAGIC energy band. It is remarkable to observe a possible softening of the spectra measured by MAGIC during the time span of a few tens of minutes.

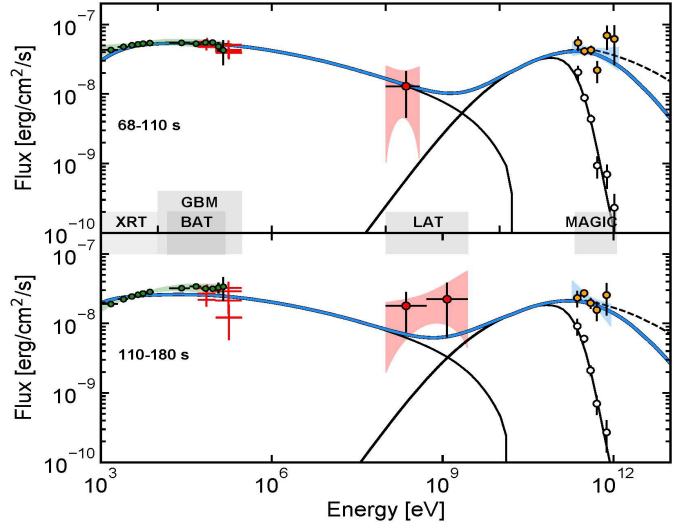

Figure 7. Modeling of the broadband spectra in the time intervals 68–110 s and 110–180 s. Thick blue curve, modeling of the multi-band data in the synchrotron and SSC afterglow scenario. Thin solid lines, synchrotron and SSC (observed spectrum) components. Dashed lines, SSC when internal γ–γ opacity is neglected. The adopted parameters for modeling can be found in [MWL]. Empty circles show the observed MAGIC spectrum. Contour regions and data points are as in Fig. 6. Image taken from [25].

An example of the theoretical modeling is shown in Fig. 7 (blue solid curve; see [25]). The dashed line shows the SSC spectrum when internal absorption is neglected. The thin solid line shows the model spectrum including EBL attenuation, in comparison to the MAGIC observations (empty circles).
Our model indicates that comparable amount of radiated energy has been channeled through the prompt and the following afterglow phases. Also the radiated power in the synchrotron and SSC components are similar.

**6. Observation of GRB 180720B by the H.E.S.S. IACT**

Four months after the detection of GRB 190114C by MAGIC and two dozen other instruments, published in the ATels and the GCN circulars in mid January 2019, the H.E.S.S. collaboration reported on a detection of a gamma-ray signal from GRB 180720B at the CTA symposium in Bologna on May 8th 2019. The H.E.S.S. observations were performed 10 hours after the onset of the burst for ~4 hours in the zenith angle range of 40°-25° degrees. The signal strength of 5 σ has been measured in the energy interval 100-440 GeV. The redshift of the GRB is z=0.653. It is remarkable that the signal from GRB 180720B has been measured so many hours after the onset of the burst, obviously deep in the afterglow phase of the GRB. It came as a real surprise that a gamma-ray signal in the range of few 100's of GeVs could be measured in such a late phase of a GRB [46].

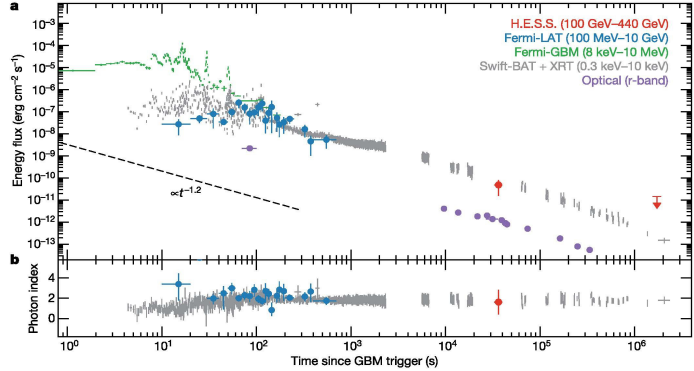

Figure 8. Multi-wavelength light curve of GRB 180720B. a, Energy-flux light-curves detected by Fermi GBM/LAT, H.E.S.S. (red arrow shows the upper limit from the second observation), Swift BAT/XRT (data extrapolated between the two instruments) and the optical r-band; the relevant colour assignment can be taken from the top-right insert. The black dashed line is for eye-guiding purposes; it shows a temporal decay with α = −1.2. b, Photon index for the spectra of Fermi-LAT, Swift and H.E.S.S. (1σ error bars). Image taken from [46].

### 7. Summary

On the night January 14/15 2019 MAGIC followed an alert from the Swift satellite mission and observed the GRB 190114C, starting from the first minute after the onset of the burst, for the next four hours. It discovered the most intense ever gamma-ray signal at VHE from GRB 190114C in the energy from 200 GeV to above 1 TeV. In the first 30 s the intensity of emission was on the level of ~130 Crab. The signal was measured in the afterglow phase although some contribution from the prompt phase cannot be excluded. This was identified as an SSC emission whose power is comparable to that of the synchrotron at lower energies. One may assume that the majority of GRBs behave similar to GRB 190114C. This is supported by the fact that most parameters inferred from our model fall within the range of those inferred from past afterglow studies. One may conclude that researchers in the past missed a significant amount of released energy.

Taking as the basis our measurement and the modeling work we anticipate that the TeV emission shall be a rather common process. This anticipation found its confirmation by the report of H.E.S.S. on the detection of GRB 180720B.

The model developed by the MAGIC collaboration, which is based on multi-wavelength data from about two dozen space-born and ground-based instruments, allowed us to reveal the second peak in the SED of GRB 190114C at the (sub)-TeV energies. We assume that from now on the GRB afterglows will regularly manifest themselves with the appearance of the second peak in their SED. Such measurements will help us to significantly extend our assessment and strongly improve understanding of the complex phenomena dubbed as GRB.

What concerns the measurement of a TeV signal from GRBs, we anticipate that as long as the energetic bursts will happen at relatively low redshifts (no extreme absorption by the EBL), the current ground-based TeV instruments shall be able to measure these, at least in the afterglow phase, even after many hours from the onset of the burst.

## 8. Afterword

The frequent question that in the past year we asked ourselves is why it took so long, say 15 years, to measure a gamma-ray signal from a GRB at TeV? The first answer is that these happen not so frequently at relatively low red-shifts (EBL absorption is strong). The second answer is that the probability is higher that these will happen at large zenith angles (i.e. at energies much higher than the threshold of a given ground-based instrument); one needs to correspondingly plan the reaction of instruments to alerts. The third reason is that one needs to observe a GRB at any, even at barely acceptable outdoor lighting conditions, once it is relevant; this further enhances the chance probability. And the last but not least answer is that one needed to show for the first time that in fact, it is do possible to measure a (sub)-TeV signal from a GRB and this is just what MAGIC did. The good thing is that the GRBs will become regular observational targets at (sub)-TeVs and the successful measurements will soon provide wealth of data allowing us to find clues to many puzzling questions about these monstrous explosions.

## References

[1] Klebesadel R.W., Strong I.B., and Olson R.A., *Ap.J.* 182, L85 (1973).

[2] Mészáros, P., Theories of gamma-ray bursts. *Annu. Rev. Astron. Astrophys*. 40, 137–169 (2002).

[3] Piran, T., The physics of gamma-ray bursts. *Rev. Mod. Phys*. 76, 1143–1210 (2005).

[4] Gehrels, N. & Mészáros, P., Gamma-ray bursts, *Science* 337, 932–936 (2012).

[5] Kumar, P. & Zhang, B., The physics of gamma-ray bursts & relativistic jets. *Phys. Rep.* 561, 1–109 (2015).

[6] Costa, E., et al., Discovery of the X-Ray Afterglow of the Gamma-Ray Burst of February 28 1997, *Nature,* v. 387, Issue 6635, pp. 783-785 (1997).

[7] Woosley, S.E., Bloom, J.S., The Supernova Gamma-Ray Burst Connection, *Ann. Rev. A&A*, v. 44, Issue 1, pp.507-556 (2006).

[8] Abbott, B.P., et al., On the Progenitor of Binary Neutron Star Merger GW170817, *ApJL*, 850, L40, (2017).

[9] Ackermann, M. et al., Fermi-LAT observations of the gamma-ray burst GRB 130427A. *Science* 343, 42–47 (2014).

[10] Piran, T. & Nakar, E., On the external shock synchrotron model for gamma-ray bursts' GeV emission. *ApJL* 718, 63–67 (2010).

[11] Inoue, S. et al., Gamma-ray burst science in the era of the Cherenkov Telescope Array, *Astropart. Phys*. 43, 252–275 (2013).

[12] Nava, L., High-energy emission from gamma-ray bursts. *Int. J. Mod. Phys. D* 27, 1842003 (2018).

[13] Mészáros, P., Rees, M. J. & Papathanassiou, H., Spectral properties of blast-wave models of gamma-ray burst sources. *ApJ* 432, 181–193 (1994).

[14] Zhang, B. & Mészáros, P., High-energy spectral components in gamma-ray burst afterglows. *ApJ* 559, 110–122 (2001).

[15] Beniamini, P., Nava, L., Duran, R. B. & Piran, T., Energies of GRB blast waves and prompt efficiencies as implied by modelling of X-ray and GeV afterglows. *MNRAS* 454, 1073–1085 (2015).

[16] Derishev, E.V., Piran, T., Particle acceleration, magnetization and radiation in relativistic shocks, *MNRAS* 460, Issue 2, 2036-2049 (2016).

[17] Derishev, E.V., Piran, T., The physical conditions of the afterglow implied by MAGIC's sub-TeV observations of GRB 190114C, *ApJL*, 880, L27 (10pp), 2019 August 1

[18] Aharonian, F.A., Very High Energy Cosmic Gamma Radiation: A Crucial Window on the Extreme Universe, World Scientific Publishing Co Pte Ltd (1762) (2004).

[19] Dermer, C.D., Menon, G., High Energy Radiation from Black Holes: Gamma Rays, Cosmic Rays, and Neutrinos, Princeton University Press, November 2009.

[20] Gropp, J. D., GRB 190114C, Swift detection of a very bright burst with a bright optical counterpart. *GCN Circulars 23688* https://gcn.gsfc.nasa.gov/gcn3/23688.gcn3 (2019).

[21] Hamburg, R., GRB 190114C, Fermi GBM detection. *GCN Circulars 23707* https://gcn.gsfc.nasa.gov/gcn3/23707.gcn3 (2019).

[22] Krimm, H. A. et al., GRB 190114C: Swift-BAT refined analysis. *GCN Circulars 23724* https://gcn.gsfc.nasa.gov/gcn3/23724.gcn3 (2019).

[23] Mirzoyan, R., et al., First time detection of a GRB at sub-TeV energies; MAGIC detects the GRB 190114C. *The Astronomer's Telegram* 12390 http://www.astronomerstelegram.org/?read=12390 (2019).

[24] Mirzoyan, R. et al., MAGIC detects the GRB 190114C in the TeV energy domain. *GCN Circular* 23701 https://gcn.gsfc.nasa.gov/gcn3/23701.gcn3 (2019).

[25] MAGIC Collaboration et al., Observation of inverse Compton emission from a long γ-ray burst. *Nature,* v. 575, 21$^{st}$ November 2019, p. 459.

[26] Acciari, V.A., Ansoldi, S., et al., Teraelectronvolt emission from the γ-ray burst GRB 190114C. *Nature,* v. 575, 21$^{st}$ November 2019, p. 458.

[27] Li, T. P. & Ma, Y. Q., Analysis methods for results in gamma-ray astronomy. *ApJ* 272, 317–324 (1983).

[28] Domínguez, A. et al., Extragalactic background light inferred from AEGIS galaxy-SED-type fractions. *MNRAS* 410, 2556–2578 (2011).

[29] Aleksić, J. et al., The major upgrade of the MAGIC telescopes, part I: the hardware improvements and the commissioning of the system. *Astropart. Phys*. 72, 61–75 (2016).

[30] Aleksić, J. et al. The major upgrade of the MAGIC telescopes, part II: a performance study using observations of the Crab Nebula. *Astropart. Phys*. 72, 76–94 (2016).

[31] Bradbury, S., et al., A Project for a 17 m Diameter Imaging Cherenkov Telescope, *Proc. 24$^{th}$ ICRC*, ed. N. Iucci, E. Lamanna, v. 1, p. 1051 (1995).

[32] Weekes, T.C., TeV Gamma-Ray Astronomy, *Nucl. Instr. Meth. Phys. Res. A* 264 (1988) 55-63.

[33] Mirzoyan, R., Constraints in Reducing the threshold of IACTs, *Proc. "Towards A Major Atmospheric Cherenkov Detector - V" workshop*, ed. O.C. De Yager, Kruger National Park, South Africa, 8-11 August, 298-314 (1997)

[34] Inoue, S., et al., *35th ICRC*, Busan, South Korea, 2017.

[35] Padilla, L., et al., Search for gamma-ray bursts above 20 TeV with the HEGRA AIROBICC Cherenkov array, *A&A* 337, 43–50 (1998).

[36] Atkins, R., et al., The High-Energy Gamma-Ray Fluence and Energy Spectrum of GRB 970417a From Observations With MILAGRITO, *ApJ*, 583, p. 824 (2003).

[37] Carosi, A. et al. Recent follow-up observations of GRBs in the very high energy band with the MAGIC Telescopes. *Proc. 34th ICRC (ICRC2015),* eds. Borisov, A. S. et al., 809 (POS, 2015).

[38] Aliu, E., et al., Observation of Pulsed γ-Rays Above 25 GeV from the Crab Pulsar with MAGIC, *Science* 21 Nov 2008, v. 322, Issue 5905, pp. 1221-1224.

[39] Ceribella, G., et al., MAGIC studies of the Crab Pulsar and Nebula spectra for energies above 20 GeV, *36th ICRC*, Madison, USA, PoS(ICRC2019)645.

[40] Lopez, M., et al., Detection of the Geminga pulsar with the MAGIC telescopes, *36th ICRC*, Madison, USA, PoS(ICRC2019)728.

[41] Kranich, D., Mirzoyan, R., et al., TeV y-ray observations of the Crab and Mkn 501 during moonshine and twilight, *Astropart. Phys.* 12, 65-74 (1999).

[42] Ahnen, M. L. et al., Performance of the MAGIC telescopes under moonlight. *Astropart. Phys*. 94, 29–41 (2017).

[43] Mirzoyan, R., Vovk, I., et al., Extending the observation limits of Imaging Air Cherenkov Telescopes toward horizon, *Nucl. Instr. Meth. Phys. Res. A*, https://doi.org/10.1016/j.nima.2018.11.046

[44] Acciari, V.A., et al., MAGIC very large zenith angle observations of the Crab Nebula up to 100 TeV, *submitted to A&A*.

[45] H. Abdalla, H., Adam, R., et al., A very-high-energy component deep in the γ-ray burst afterglow, *Nature,* v. 575, 21st November 2019, p.464.

[46] Galli, A., Piro, L., Prospects for detection of very high-energy emission from GRB in the context of the external shock model, *A&A* 489, 1073–1077 (2008).